\documentclass[twocolumn,showpacs,preprintnumbers,amsmath,amssymb]{revtex4}
\usepackage{graphicx}
\usepackage{epsfig}
\usepackage{dcolumn}
\usepackage{bm}

\newcommand{\kp}{K^+}
\newcommand{\kn}{K^-}
\newcommand{\kstp}{K^{*+}}

\newcommand{\ksto}{K^{*0}}

\newcommand{\be}{\begin{enumerate}}
\newcommand{\ee}{\end{enumerate}}
\newcommand{\bi}{\begin{itemize}}
\newcommand{\ei}{\end{itemize}}

\newcommand{\jpsi}{J/\psi}

\newcommand{\psip}{\psi(2S)}
\newcommand{\rar}{\rightarrow}
\newcommand{\ppp}{\pi^+\pi^-\pi^0}
\newcommand{\pip}{\pi^+}
\newcommand{\pin}{\pi^-}
\newcommand{\pio}{\pi^0}

\newcommand{\ar}{\rightarrow}
\newcommand{\rt}{\rightarrow}
\newcommand{\etal}{\it et al. \rm}
\newcommand{\uu}{\mu^+\mu^-}
\newcommand{\epen}{e^+e^-}

\newcommand{\ppj}{\pi^+\pi^-J/\psi}

\def\Journal#1&#2&#3(#4){#1{\bf #2}, #3 (#4)}

\def\etal{et al.}

\def\bec{\begin{center}}
\def\eec{\end{center}}

\begin{document}
\title{\Large \bf \boldmath Measurement of the Branching Fraction of $\jpsi\to\ppp$ }
\author{J.~Z.~Bai$^1$,        Y.~Ban$^{10}$,         J.~G.~Bian$^1$,
X.~Cai$^{1}$,         J.~F.~Chang$^1$,       H.~F.~Chen$^{16}$,    
H.~S.~Chen$^1$,       H.~X.~Chen$^{1}$,      J.~Chen$^{1}$,        
J.~C.~Chen$^1$
% ...
\\(BES Collaboration)\\ 
\vspace{0.2cm} 
$^1$ Institute of High Energy Physics, Beijing 100039, People's Republic of
     China\\
$^2$ China Center of Advanced Science and Technology, Beijing 100080,
     People's Republic of China\\
$^3$ Guangxi Normal University, Guilin 541004, People's Republic of China\\
$^4$ Henan Normal University, Xinxiang 453002, People's Republic of China\\
$^5$ Huazhong Normal University, Wuhan 430079, People's Republic of China\\
$^6$ Hunan University, Changsha 410082, People's Republic of China\\                                                  
$^7$ Liaoning University, Shenyang 110036, People's Republic of China\\
$^{8}$ Nanjing Normal University, Nanjing 210097, People's Republic of China\\
$^{9}$ Nankai University, Tianjin 300071, People's Republic of China\\
$^{10}$ Peking University, Beijing 100871, People's Republic of China\\
$^{11}$ Shandong University, Jinan 250100, People's Republic of China\\
$^{12}$ Shanghai Jiaotong University, Shanghai 200030, 
        People's Republic of China\\
$^{13}$ Sichuan University, Chengdu 610064,
        People's Republic of China\\                                    
$^{14}$ Tsinghua University, Beijing 100084, 
        People's Republic of China\\
$^{15}$ University of Hawaii, Honolulu, Hawaii 96822\\
$^{16}$ University of Science and Technology of China, Hefei 230026,
        People's Republic of China\\
$^{17}$ Wuhan University, Wuhan 430072, People's Republic of China\\
$^{18}$ Zhejiang University, Hangzhou 310028, People's Republic of China\\
\vspace{0.4cm}
$^{\ast}$ Visiting professor to University of Michigan, Ann Arbor, MI 48109 USA 
}

\noindent\vskip 0.2cm 
\begin{abstract}
Using 58 million $\jpsi$ and 14 million $\psip$ decays obtained by the
BESII experiment, the branching fraction of $\jpsi\to\ppp$ is
determined.  The result is $(2.10 \pm 0.12) \times 10^{-2}$, which is
significantly higher than previous measurements.
\end{abstract}

\maketitle 

\section{Introduction}   \label{introd} 
Decays of the $\jpsi$ provide an excellent source of events with which
to study light hadron spectroscopy and search for glueballs, hybrids,
and exotic states.  Since the discovery of the
$\jpsi$ at Brookhaven~\cite{aube} and SLAC~\cite{augu} in 1974, more
than one hundred exclusive decay modes of the $\jpsi$ have been
reported.  Recently, $5.8\times 10^{7}$ $\jpsi$ events and $1.4\times
10^{7}$ $\psip$ events have been obtained with the upgraded Beijing
Spectrometer (BESII), and these samples offer a unique opportunity to
measure precisely the branching fractions of $\jpsi$ decays.

The largest $J/\psi$ decay involving hadronic resonances is $J/\psi
 \rt \rho(770) \pi$.  Its branching fraction has been reported by many
 experimental
 groups~\cite{marki,brau,desy1,pluto,dasp,markii,markiii,besi}
 assuming all $\pi^+ \pi^- \pi^0$ final states come from $ \rho(770)
 \pi$.
The precision of these measurements varies from $13\%$ to 25\%.
In this paper, we present two independent measurements of this branching
fraction using $\jpsi$ and $\psip$
decays.
The first is an absolute measurement based on $\jpsi\ar\ppp$ directly.
The second, in which many of the systematic errors cancel out,
 is a relative measurement obtained from a comparison
of the rates for $\jpsi\ar\ppp$ and $\jpsi\ar\uu$, using $J/\psi$
events produced via $\psip\ar\ppj$.

\section{The BES Detector}  \label{BESD} 

The upgraded BESII detector operates
at the Beijing Electron-Positron Collider (BEPC); it is a large solid-angle
magnetic spectrometer that is described in detail in Ref.~\cite{besii}.
The momentum of charged particles is determined by a
40-layer cylindrical main drift chamber (MDC) which has a momentum
resolution of 
 $\sigma_{p}/p = 1.78\%\sqrt{1+p^2}$, where $p$ is in units of GeV/c.
Particle identification is accomplished using specific ionization ($dE/dx$)
measurements in the drift chamber and time-of-flight (TOF) information in 
a barrel-like array of 48 scintillation counters. The $dE/dx$ resolution
is $\sigma_{dE/dx}=8.0\%$; the TOF resolution for Bhabha events is
$\sigma_{TOF}=180 $ ps.
Radially outside the time-of-flight counters is a 12-radiation-length
barrel shower counter (BSC) comprised of gas 
proportional tubes interleaved with lead sheets. The BSC measures
the energies and directions of photons with resolutions
of $\sigma_{E}/E\simeq21\%\sqrt{E({\rm GeV})}$, $\sigma_{\phi}=7.9$ mrad, and
$\sigma_{z}=2.3$ cm. The iron flux return of the magnet is instrumented
with three double layers of counters (MUC) that are used to identify muons.

In the analysis, a GEANT3 based Monte Carlo program (SIMBES) with detailed 
consideration of detector performance (such as dead electronic channels)
is used. The consistency between data and Monte Carlo has been
checked in many high purity physics channels, and the agreement is
reasonable.


\section{General Criteria}  \label{gen_sel} 
\subsection{Charged Track Selection}  
\label{method}
Each charged track, reconstructed using hits in the MDC, must (1) have a
good helix fit, in order to ensure a correct error matrix in the
kinematic fit; (2) originate from the interaction region,
$\sqrt{V_x^2+V_y^2}<2$ cm and $|V_z|<20$ cm, where $V_x$, $V_y$, and
$V_z$ are the x, y, and z coordinates of the point of closest approach
of the track to the beam axis; (3) have a transverse momentum greater
than 60 MeV/c; and (4) have $|\cos \theta|\leq 0.8$, where $\theta$ is
the polar angle of the track.

\subsection{Photon Selection}
A neutral cluster in the BSC is assumed to be a photon candidate if the
following requirements are satisfied: (1) the energy deposited in the
BSC is greater than 0.06 GeV; (2) the total number of layers
with deposited energy is greater than one; (3) the angle
between the direction of photon emission and the direction of shower
development is less than $30^{\circ}$; and (4) the angle
between the photon and the nearest charged track is
greater than $15^{\circ}$.  If the angle between two neutral clusters
is less than $10^{\circ}$ and their $\gamma \gamma$ invariant mass is
less than 0.05 GeV/c$^2$, they are combined with the cluster with the
largest energy being used for the direction and energy of the combined
cluster in the kinematic fit.

\section{\boldmath Absolute Measurement of $\jpsi\rar\ppp$ Decays}

\subsection{Event Selection}
Events with two oppositely charged tracks and at least two good
photons are selected for further analysis. No charged particle
identification is required. A 5-constraint (5C) kinematic fit is made
under the $\pi^+\pi^-\gamma\gamma$ hypothesis with the invariant mass of the
two photons being constrained to the $\pi^0$ mass. 
If the number of the selected photons is larger than two, the fit is
repeated using all permutation of the photons. For events with a good fit,
the two photon combination with the minimum fit  $\chi^{2}_{\ppp}$ 
is selected, and its value is required to be less than 15.


To select a clean sample, the following criteria are applied
to the remaining events:

\begin{itemize}

\item To reject the main background events from
$\jpsi\ar K^+K^-\pi^0$,
a 5C kinematic fit for $\jpsi\ar K^+K^-\pi^0$ is performed, and
$\chi^{2}_{\ppp}<\chi^{2}_{K^+K^-\pi^0}$ is required.
Fig.~\ref{chisq} shows the scatter plot of $\chi^{2}_{\ppp}$ versus
$\chi^{2}_{K^+K^-\pi^0}$.

\begin{figure}[htbp]
\centerline{\epsfig{figure=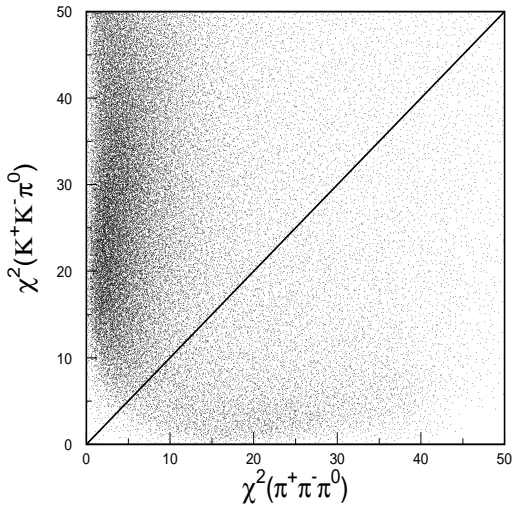,height=7.cm,width=0.4\textwidth}}
\caption{\label{chisq} Plot of $\chi^{2}_{\ppp}$ versus
 $\chi^{2}_{K^+K^-\pi^0}$ for candidate $\pi^+\pi^-\pi^0$ events.
The solid line corresponds to $\chi^{2}_{\ppp}$= $\chi^{2}_{K^+K^-\pi^0}$. }
\end{figure}

\item  Background events  from 
$\gamma$ conversions ($\gamma \rt e^+ e^-$) are eliminated by requiring the  
 angle between the two charged tracks, $\theta_{\pip\pin}$, to be greater than
$10^{\circ}$.



\item Radiative events, for example $\jpsi\to\gamma\eta^{\prime}$, are
removed by the requirement $|\cos\theta_{\gamma}|<0.98$, where $\theta
_{\gamma}$ is the angle of the $\gamma$ in the $\pi^0$ rest frame.


\item Contamination from $\jpsi\ar(\gamma) \epen$ is eliminated by the
requirement that the sum of the deposited energies of the two charged
pions in the BSC is less than 2 GeV.  Fig. \ref{e1e2} shows the
scatter plot of $E_{sc}^+$ versus $E_{sc}^-$, where $E_{sc}^+$ and
$E_{sc}^-$ are the deposited energies of the $\pip$ and $\pin$ in the
BSC, respectively. This criteria has almost no effect on $\ppp$
events.


\begin{figure}[htbp]
\centerline{\epsfig{figure=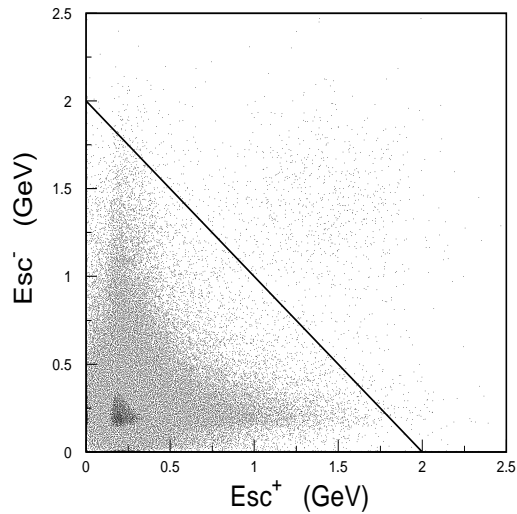,height=7.cm,width=0.4\textwidth}}
\caption{\label{e1e2}
  Plot of $E_{sc}^+$ versus $E_{sc}^-$; the solid
  line is for $E_{sc}^+$+ $E_{sc}^-$ = 2 GeV.}
\end{figure}
\end{itemize}

After the above requirements, 219691 $\ppp$ candidates are selected.
Remaining backgrounds are evaluated using two different Monte Carlo
simulations. In the first, specific background channels, shown in
Table \ref{bkg}, are generated.  The total background from these
channels in the selected $\ppp$ events is determined to be less than
1\%.  The second simulation uses 30 million inclusive $\jpsi$ MC
events generated with the LUND model~\cite{LUND}. After normalizing
the selected background events to 58 million $\jpsi$ events, 3799
background events are obtained, yielding a contamination of 1.7\%. In
this paper, the latter background estimate is used to correct the
branching fraction, giving a correction factor of $(98.3\pm 1.7)$\%.

\begin{table}[htpb]
\caption{Background contributions from different decay channels. Here
$N_{bkg}$ is the number of events generated, and $N_{bkg}^{norm}$ is
the number of background events selected, normalized by the branching
fractions quoted in Ref. \cite{pdg2002}.  }
\begin{center}
\begin{tabular}{l|c|c}
\hline
\hline
Decay Channel& $N_{bkg}$ & $N^{norm}_{bkg}$\\
\hline
$\jpsi\ar \kstp\kn+c.c.$ $(\kp\kn\pi^0)$ &100,000 &773 \\
$\jpsi\ar \kstp\kn+c.c.$ $(K K^0_S\pi)$ &50,000 &153 \\
$\jpsi\ar \ksto\bar{K^0}+c.c.$ $(K K^0_S \pi)$ &50,000 &129\\
$\jpsi\ar \gamma\eta^{\prime}$ $(\gamma\gamma\rho)$ & 100,000 & 158 \\
\hline
\hline
\end{tabular}
\label{bkg}
\end{center}
\end{table}

\begin{figure}[htbp]
\centerline{\epsfig{figure=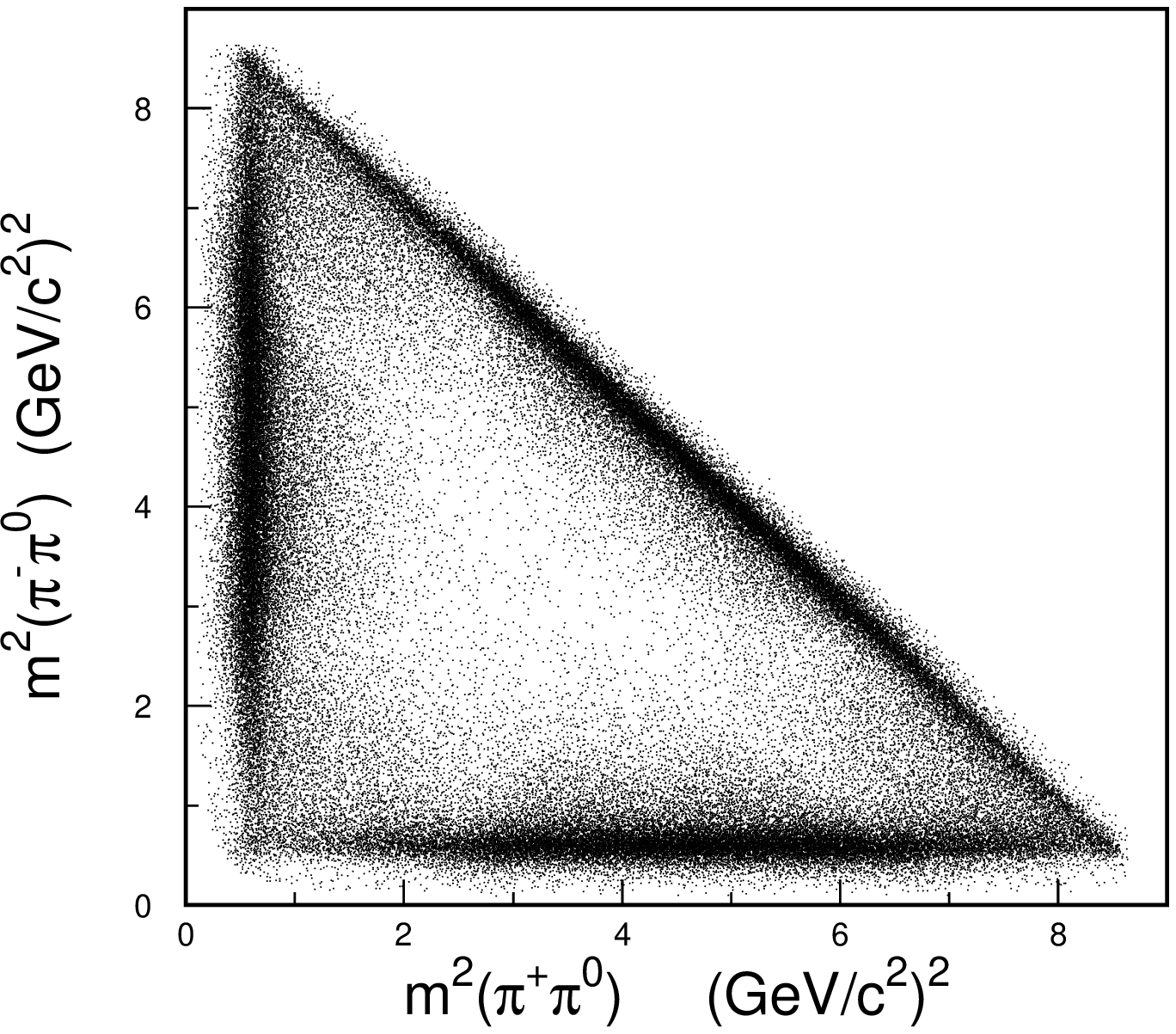,height=6.8cm,width=0.4\textwidth}}
\caption{\label{dalitz}
  The Dalitz plot for $\jpsi\to\ppp$.}
\end{figure}

\begin{figure}[htbp]
\centerline{\epsfig{figure=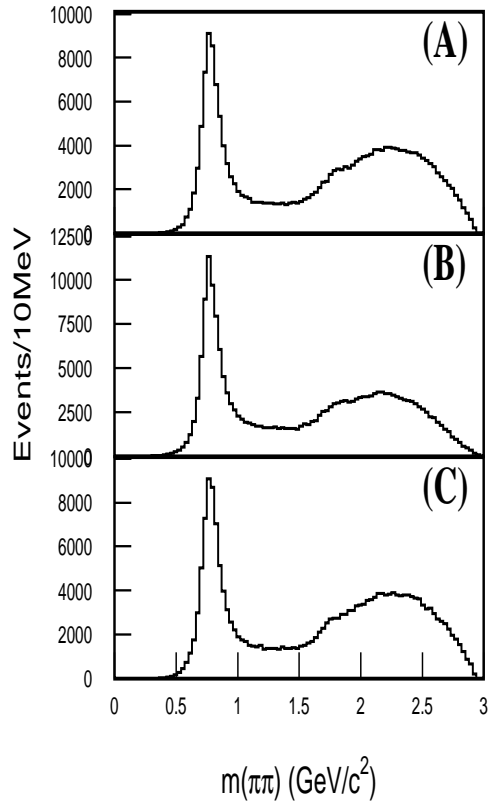,height=11.5cm,width=0.4\textwidth}}
\caption{\label{hist}
  The distributions of invariant mass of two pions for (A)
  $m_{\pip\pio}$, (B) $m_{\pip\pin}$,  and (C) $m_{\pin\pio}$.}
\end{figure}

The Dalitz plot of $m_{\pi^+\pi^0}$ versus $m_{\pi^-\pi^0}$ is shown in
Fig. \ref{dalitz}. Three  bands are clearly visible in the plot, which 
 correspond to $\jpsi\to\rho^+\pi^-$,
$\jpsi\to\rho^0\pi^0$, and 
$\jpsi\to\rho^-\pi^+$. The corresponding histograms of
$m_{\pip\pio}$, $m_{\pip\pin}$, and $m_{\pin\pio}$ are shown in
Fig. \ref{hist}.
From the Dalitz plot (Fig. \ref{dalitz}), we see that $\jpsi\ar\ppp$ is
strongly dominated by $\rho\pi$. Therefore, the detection efficiency
is determined using the RHOPI~\cite{rhopi} generator with SIMBES and
is found to be 17.83\%.
Monte Carlo simulation using other generators to represent the structure
in the Dalitz plot, provide very similar detection efficiencies.
 
\subsection{Systematic Error Analysis} \label{J-sys}
In this analysis, the  systematic error on the branching fraction comes mainly
from the following sources:
\begin{itemize}
\item MDC tracking
 
The MDC tracking efficiency has been measured using channels like
$e^+e^-\rar(\gamma)e^+e^-$, $e^+e^-\rar(\gamma)\mu^+\mu^-$,
$\jpsi\ar\Lambda\bar{\Lambda}$, and $\psi(2S)\ar\pip\pin\jpsi,$
$\jpsi\ar\mu^+\mu^-$. It is found that the Monte Carlo simulation
agrees with data within 1-2\% for each charged track. The systematic
error on the tracking efficiency for the channel
of interest is taken as 4\%.
\item Photon detection efficiency

 The photon detection efficiency is studied with $\jpsi\ar\rho^0\pio$
events. Events with two oppositely charged tracks and at least one
photon are selected. The two charged tracks are required to be
identified as pions using particle identification. A 2C kinematic fit
is made under the hypothesis $\pi^+\pi^-\gamma\gamma_{missing}$, where
$\gamma_{missing}$ is a missing photon and the
$\gamma\gamma_{missing}$ invariant mass is constrained to the $\pi^0$
mass. The combination with the smallest $\chi^2$ is selected and
is required to satisfy $\chi^2<10$, as well as be less than the
$\chi^{2}$ of the 2C kinematic fit with the two charged tracks assumed
to be kaons.  Events with 0.62 GeV/c$^2 < m_{\pip\pin}<0.92$
GeV/c$^2$, $m_{\pin\pio}>1.2$ GeV/c$^2$, and $m_{\pip\pio}>1.2$
GeV/c$^2$ are selected as $\rho^0$ candidates. The ``missing" photon's
energy distribution is large enough to cover the case
of $\rho^{\pm}\pi^{\mp}$, so it is
used to study the photon detection efficiency. The same analysis is
performed with Monte Carlo events.
The gamma detection efficiency for data is in good agreement with that from Monte
Carlo. The difference between them is about 2\% for each photon which is 
taken as the systematic error. 


\item Kinematic fit and other criteria

To estimate the systematic error from the 5C kinematic fit, we select
a clean $\rho\pi$ sample without the kinematic fit. 
 Events with two oppositely charged tracks and two good
photons are selected.  The charged tracks must be identified as
pions. The direction of ${\bf P_{miss} }$, where ${\bf P_{miss} }$ is
the missing momentum determined using the charged tracks, is regarded
as the direction of the $\pio$ and used to calculate the invariant
mass of the two photons, which is required to be less than 0.2
GeV/c$^2$. A variable $U_{miss}=E_{miss}-|{\bf P_{miss}}|$ is defined,
where $E_{miss}$ is the missing energy of the two charged tracks which is
calculated assuming the charged tracks are pions.  $U_{miss}$
is required to be less than zero to select a clean sample.

 A 5C kinematic fit is done on the candidates.
The same analysis is also
performed with Monte Carlo events.  By comparing the number of events
with and without a good 5C kinematic fit, the efficiencies for
$\chi^{2}_{\pip\pin\pio}<15$ are measured to be $76.5\%$ and
$79.8\%$ for real data and Monte Carlo simulation, respectively. The
difference between them is 4.1\% and a correction factor, 1.041, is
obtained, and the systematic
error on this correction is taken as 4.1\%. 



To estimate the systematic error from the
$\chi^{2}_{\pip\pin\pio}<\chi^{2}_{\kp\kn\pio}$ requirement, we
selected events where both charged tracks are identified as pions
using particle identification. The branching fraction is then obtained
with all the selection criteria described above. The difference
between this result and that calculated without the
$\chi^{2}_{\pip\pin\pio}<\chi^{2}_{\kp\kn\pio}$ requirement is
regarded as the systematic error caused by this criteria. In fact, the
difference between them is very small, and the error caused by this
criteria is less than 1\%.

For the requirements $\theta_{\pip\pin}>10^{\circ}$ and $|\cos\theta_{\gamma}|<0.98$,
just a few events are excluded by these selection criteria; the
systematic error for them can be ignored.  From the scatter plot
shown in Fig. \ref{e1e2}, the requirement on the deposited energy of
two charged pions has almost no effect on the $\ppp$ candidates, so
the systematic error from this selection criteria is also neglected.
The total systematic error from the kinematic fit and other criteria
discussed in this section is 4.2\%, which is the sum of these errors added in quadrature.

\item Uncertainty of the hadronic model

Different simulation models for the hadronic interaction give
different efficiencies, leading to different branching fractions.
In this analysis, two models, FLUKA~\cite{FLUKA} and GCALOR~\cite{GCALOR},
are used in the simulation of hadronic interactions in SIMBES.
The difference between the detection
efficiencies from them is about 1.7\%, which is
regarded as the systematic error. 

\item Uncertainty of background

Above we estimated backgrounds for several
possible decay channels. Given the uncertainties of the branching fractions
of background channels and possible unknown decay modes of $\jpsi$, we
estimate the uncertainty of the background is less than 3\%.


\end{itemize}

The contributions from all sources are listed in Table
\ref{toterr}.  The systematic errors caused by Monte Carlo statistics
and the error in the number of $\jpsi$ events are also listed. The
total systematic error in Table \ref{toterr} is the sum of them added in
quadrature.

\begin{table}[h] \centering
\caption{\label{toterr}Summary of correction factors $f_c$ and systematic errors}

\begin{tabular}{l|c|c}  
\hline
\hline 
Sources & $f_c$ &Systematic error (\%)\\
\hline
MDC tracking &  &  4\\ 

Photon  efficiency & & 4 \\

Kinematic fit &1.041  &4.2 \\

Hadronic model &  & $\sim 3$\\

Backgrounds &0.983 & 1.7 \\

MC statistics & & 0.4\\

Number of $\jpsi$ events & & 4.7\\
\hline
Total  & 1.023 & 9.2 \\
 \hline
\hline
\end{tabular}
\end{table}
\subsection{\boldmath Branching Fraction of $\jpsi\ar\ppp$ }  \label{selec}
For the decay of $\jpsi\ar\ppp$, the branching fraction is obtained
with the following formula

\begin{equation}
Br(\jpsi\ar\ppp)={{N^{obs}_{\ppp}}\over{N_{\jpsi}\cdot\epsilon}}\cdot f_c
\end{equation} 
where ${N^{obs}_{\ppp}}$ is the observed number of $\ppp$ events,
$\epsilon$ is the detection efficiency obtained from the MC
simulation, $N_{\jpsi}$ is the total number of $\jpsi$ events,
$(57.7 \pm 2.7)\times 10^{6}$, which is determined from the
number of inclusive 4-prong hadrons~\cite{hepnp}; and
$f_c$ is the kinematic fit and background
contamination correction factor.

With the above formula, the branching fraction of $\jpsi\ar\ppp$ is

\begin{center}
$Br(\jpsi\ar\ppp)=(21.84\pm0.05\pm 2.01)\times 10^{-3}$
\end{center}
where the first error is statistical and the second systematic.

\section{\boldmath Relative Measurement of $\jpsi\ar\ppp$}
 The relative measurement is based on a sample of 14 million $\psi(2S)$ events. 
The $\psi(2S)$ is a copious source of
$J/\psi$ decays: the branching fraction of $\psi(2S)\rightarrow\pi^+\pi^-
J/\psi$ is the largest single $\psi(2S)$ decay channel. Therefore,
we can determine the branching fraction of $J/\psi\rightarrow\pi^+\pi^-\pi^0$
from a comparison of the following two processes:
\begin{eqnarray*}
\psi(2S)\rightarrow\pi^+\pi^- &J/\psi& \\
                     &\hookrightarrow &\pi^+\pi^-\pi^0 ~~~~~~~~~~~~~~~~(I) \\
 {\rm and}                 &\hookrightarrow & \mu^+\mu^- ~~~~~~~~~~~~~~~~~(II)
\end{eqnarray*}
The branching fraction is determined from the relation:
\begin{equation}
\begin{array}{ccl}
B(J/\psi\rightarrow\pi^+\pi^-\pi^0)
&=&\frac{N_I^{obs}}{N_{II}^{obs}}\cdot
 \frac{\epsilon_{II}}{\epsilon_I}\cdot
  {B(J/\psi\rightarrow \mu^+\mu^-)},
\end{array} \label{formula2}
\end{equation}
where $N_I^{obs}$ and $N_{II}^{obs}$ are the observed numbers of
events for processes I and II, and $\epsilon_{I}$ and $\epsilon_{II}$
are the respective acceptances.  The branching fraction for the
leptonic decay $J/\psi\rightarrow \mu^+\mu^-$, $B(J/\psi\rightarrow
\mu^+\mu^-)= (5.88\pm0.10)\%$, is obtained from the Particle Data
Group (PDG)~\cite{pdg2002}.  Using the relative measurement, many
systematic errors, for instance, the errors of the total
number of $\psip$ events, the branching fractions of
$\psip\rar\pi^+\pi^-\jpsi$, and the efficiency for
$\psip\rar\pi^+\pi^-\jpsi$, etc, mostly cancel. Therefore, the
precision of the branching fraction $\jpsi\rar\pi^+\pi^-\pi^0$ from
the relative measurement is
comparable with that of the direct $\jpsi$ decay, as we will see
later, although the size of $\psip$ sample is smaller than the $\jpsi$
sample.

\subsection{Event Selection}

Candidate events for $\psi(2S)\rightarrow \pi^+\pi^- J/\psi$,
$J/\psi\rightarrow\mu^+\mu^-$ or $\pi^+\pi^-\pi^0$
are required to have four charged tracks with
total charge zero. Each track is required to satisfy the general
criteria described in Section~\ref{gen_sel}.
  For both processes I and II, we require at least one pair of
oppositely charged candidate pion tracks that each satisfy the following
criteria:
\begin{itemize}
\item $p_{\pi}<0.5$ GeV/c, where $p_{\pi}$ is the pion momentum.
\item $\cos\theta_{\pi\pi}<0.9$, where $\theta_{\pi\pi}$ is the
laboratory angle between the $\pi^+$ and $\pi^-$. This requirement is
used to eliminate contamination from misidentified $e^+e^-$ pairs from
$\gamma$ conversions.
\end{itemize}
 The invariant mass recoiling against the candidate $\pi^+\pi^-$ pair,
$m_{recoil}^{\pi^+\pi^-}=
[(m_{\psi(2S)}-E_{\pi^+}-E_{\pi^-})^2-
({p}_{\pi^+}+{ p}_{\pi^-})^2]^{1/2}$,
is required to be in the range
$3.0\leq m_{recoil}^{\pi^+\pi^-}\leq 3.2$ GeV/c$^2$.

For process I, candidate events are required
 to  satisfy the following additional criteria: 
\begin{itemize}
\item  All four charged tracks are assumed to be $\pi^{\pm}$
  directly, and no particle identification is required.
\item  The number of photon candidates must be equal to or greater than two.
\item  A 5C kinematic fit is performed for each
$\psi(2S)\rightarrow\pi^+\pi^-\pi^+\pi^-\pi^0$ candidate event, and
 the event probability given by the fit must be greater than 0.01 and
 greater than that of $\psi(2S)\rightarrow\pi^+\pi^-K^+K^-\pi^0$.
\item Remaining background from $J/\psi$ to $e^+e^-$ and $\mu^+\mu^-$ events is removed with
  the following requirement:
  $[(\chi_e^++\chi_e^-)^2/9+(E_{sc}^++E_{sc}^++\mu^+_{id}+\mu^-_{id}-2.5)^2/3]>1$
  and $E_{sc}^++E_{sc}^-+\mu^+_{id}+\mu^-_{id}<6$, as shown in
  Fig. \ref{5pi_ep}. Here $\chi_e$ is the difference between the $dE/dx$ measured
  with the MDC and that expected for the electron hypothesis divided
  by the  $dE/dx$ resolution, $E_{sc}$ is the energy deposited in the BSC, and $\mu_{id}$ is
  the number of MUC layers with matched hits and ranges from 0 to 3.
  The contamination from $\jpsi\rightarrow e^+e^-$ or $\mu^+\mu^-$ is
  estimated to be 0.4\% from Monte Carlo simulation.

 \begin{figure}[htbp]   \centering
  \includegraphics[width=0.45\textwidth]{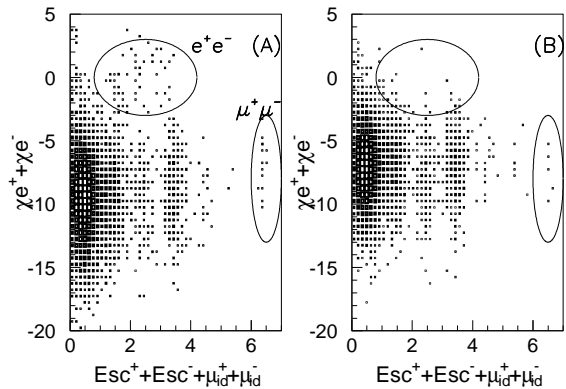}
  \caption{\label{5pi_ep}
   Scatter plots of $\chi_{e^+}+\chi_{e^-}$ versus
    $E_{sc}^++E_{sc}^-+\mu^+_{id}+\mu^-_{id}$  for (a.) data and (b.)
    MC.}
 \end{figure}

\end{itemize}

\begin{figure}[htbp]  \centering
  \includegraphics[width=0.45\textwidth]{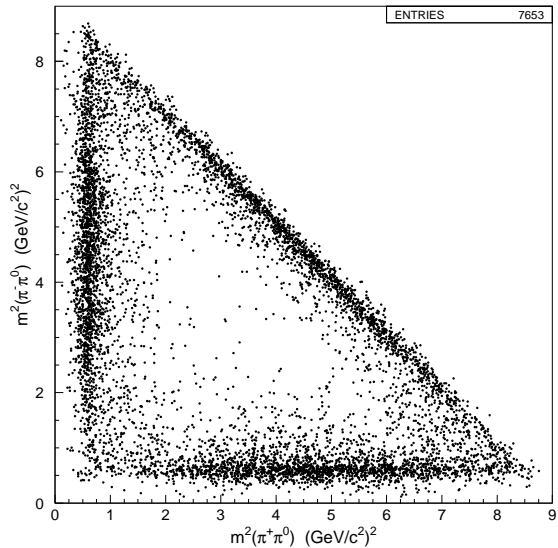}
  \caption{\label{5pi_rhopi_dalitz} Dalitz
  plot for candidate $J/\psi\rightarrow\pi^+\pi^-\pi^0$ events. Here
  the $J/\psi$ comes
  from $\psip\rightarrow \pi^+\pi^-J/\psi$ decay.}
  \end{figure}

The Dalitz plot of candidate $J/\psi\rightarrow \pi^+\pi^-\pi^0$
events is shown in Fig. \ref{5pi_rhopi_dalitz}.  The contamination
from $J/\psi\rightarrow K^*K$ is about 1.0\% and is estimated from
Monte Carlo
simulation. Those of other backgrounds (e.g. $e^+e^-$, $\mu^+\mu^-$, and
$\gamma\gamma \rho$) are much less than 1.0\%.

To reduce possible systematic bias caused by inconsistencies between
data and Monte Carlo, similar requirements are used
for $J/\psi\rightarrow\mu^+\mu^-$ candidate events (process II).

\begin{figure}[htbp]  \centering
  \includegraphics[height=7.cm,width=0.45\textwidth]{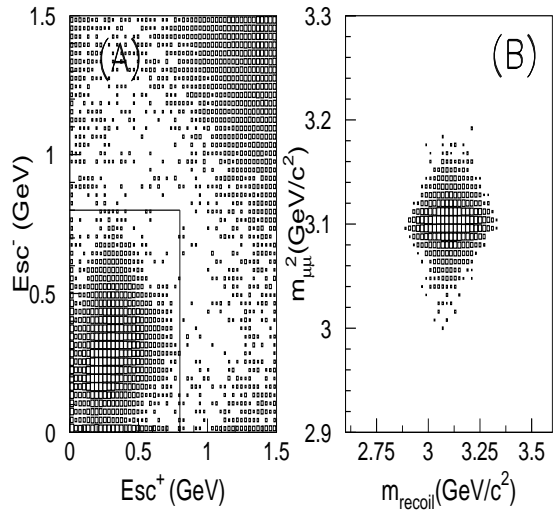}
  \caption{\label{pi2mu2_1}
   (a) Deposited energy in BSC for $\mu^+$ and $\mu^-$ and (b.)
   plot of $m_{recoil}^{\pi^+\pi^-}$ versus $m_{\mu^+\mu^-}$ for
   candidate $\psip\rightarrow\pi^+\pi^-J/\psi,J/\psi\rightarrow\mu^+\mu^-$
   events.}
 \end{figure}

\begin{itemize}
\item The two higher momentum tracks are assumed to
be $\mu^{\pm}$, and no muon identification is required.
\item  A 4C kinematic fit is performed for
$\psi(2S)\rightarrow\pi^+\pi^-\mu^+\mu^-$ candidate events, and
 the probability given by the fit must be greater than 0.01.
\item The contamination from $J/\psi\rightarrow e^+e^-$ is removed with
     the requirement $E^{\pm}_{sc} <0.8 $ GeV.
    After this, the contamination from $e^+e^-$  is
    less than $0.8\%$, estimated from 
Fig. \ref{pi2mu2_1}a,
    and $\sim 0.4\%$ from MC simulation.
\end{itemize}

Fig. \ref{pi2mu2_1}b shows the scatter plot
of $m_{recoil}^{\pi^+\pi^-}$ versus $m_{\mu^+\mu^-}$ for
$\psi(2S)\rightarrow \pi^+\pi^-J/\psi\rightarrow\pi^+\pi^-\mu^+\mu^-$
candidate events.  
MC simulations of $J/\psi$ decays to $\pi^+\pi^-$, $K^+K^-$, $p\bar{p}$
 and $\rho\pi$ indicate that background from these processes can be ignored.

\begin{figure}[htbp] \centering
  \includegraphics[width=0.45\textwidth]{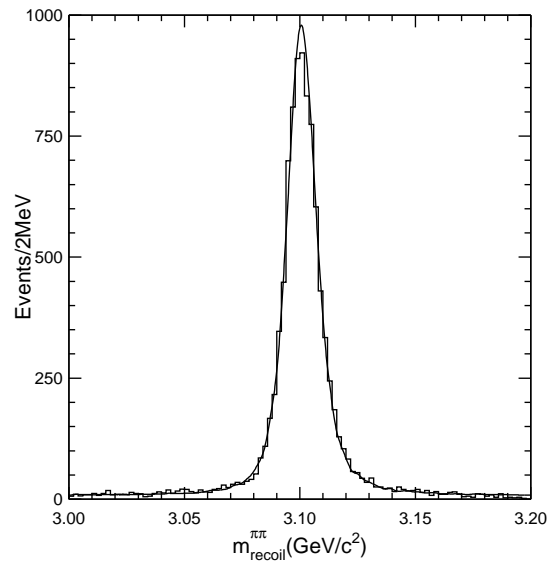}
  \caption{\label{rhopi_fit} Fitting the 
    $\pi^+\pi^-$ recoil mass of $\psi(2S)\rightarrow\pi^+\pi^-
            J/\psi$, $J/\psi\rightarrow\rho\pi$ (the histogram) with the
    $m^{\pi^+\pi^-}_{recoil}$ spectrum parameters
     obtained from $\psi(2S)\rightarrow\pi^+\pi^- J/\psi$,
     $J/\psi\rightarrow\mu^+\mu^-$ (the curve).}
 \end{figure}

By fitting the invariant mass recoiling against the $\pi^+\pi^-$ pair in
$\psip\ar\pi^+\pi^-\jpsi,$ $\jpsi\ar\uu$ decay, one obtains the
$m^{\pi^+\pi^-}_{recoil}$ spectrum, which is then used to fit the
recoil mass spectrum of the $\psip\ar\pi^+\pi^-\jpsi,\jpsi\ar\ppp$ process,
as shown in Fig~\ref{rhopi_fit}.  We obtain
$\frac{N^{obs}_{I}}{N_{II}^{obs}}=0.102\pm0.001$ and use the same
procedure on simulated data to determine
$\frac{\epsilon_I}{\epsilon_{II}}=0.286\pm0.003$, so the ratio of
$B(\jpsi\to\ppp)$ to $B(\jpsi\to\uu)$ is $(35.7\pm0.5)\%$.  Here the
errors are the statistical uncertainties combined with the uncertainties
in the fitting procedure.

\subsection{Systematic Error Analysis}

Systematic errors come from background uncertainties, the uncertainties
of $B(\jpsi\ar\uu)$ and $B(\pi^0\ar\gamma\gamma)$, and imperfections in the Monte Carlo simulations.

 Since similar requirements are used for processes I and II, 
many systematic errors  cancel out. For instance, the uncertainty in
the selection of the $\pi^+\pi^-$ pair recoiling against the $J/\psi$
will not contribute to the 
systematic error. Other uncertainties are treated in the following:
\begin{itemize}
\item MDC tracking

  This systematic error is caused by differences between MDC tracking
  efficiencies for data and Monte Carlo simulation.  Since the Monte
  Carlo simulation agrees with data within 1 to 2\% for each charged
  track, this systematic error is less than 2.0\%.

%

\item Photon detection efficiency

   Two photons are involved in process I and no photons
in process II.  The uncertainty of photon selection is about 4\%
according to the study described in Section ~\ref{J-sys}.

\item Uncertainty of the Hadronic model

Since $\jpsi\to\pi^+\pi^-\pi^0$ is strongly dominated by the
  $\rho(770)\pi$ dynamics and the contribution of the excited rho
  states is still unknown, a $\psi(2S)\to\pi^+\pi^-\jpsi$,
  $\jpsi\to\rho\pi$ simulation is used to obtain the detection efficiency
  ($\epsilon_{I}$). The effect of the excited rho states is estimated
  to be about 1.0\%.

  The difference found from different models of the hadronic
interaction (GCOLAR and FLUKA) is 1.2\%.

\item Kinematic fit

  A kinematic fit is performed for both processes I and II, and the
probability given by the fit is required to be greater than 0.01. Since the
kinematic fit depends on the error matrix from track fitting, a
systematic error of 1.5\% is estimated from the difference of the error matrix
for data and Monte Carlo simulation. In addition, 
$\chi^2(\pi^+\pi^-\ppp)<\chi^2(\pi^+\pi^-K^+K^-\pi^0)$ is used for
process I, which causes a correction factor of $(1.2\pm0.5)\%$, determined 
from an analysis similar to that described in Section~\ref{J-sys}.

\item Uncertainty of background

Above we estimated posssible backgrounds for processes I and II, and
contaminations of about 2.0\% and 0.4\% were obtained, respectively.
We have also used a sample of 14 million inclusive $\psip$ MC events
generated with the LUND model~\cite{LUND} to estimate
the contribution of background for process I, and the contamination
is found to be less than 3.0\%; a background correction factor
$(98.4\pm1.5)\%$ is used.

 Other requirements, such as those to remove $\jpsi\ar \epen$ or $\uu$ for process I
 and to remove $\jpsi\ar \epen$ for process II, 
 have very high efficiencies ($\sim100\%$). Their systematic error
 contributions are ignored.

\end{itemize}

  Table~\ref{TSystematicError} summarizes all
systematic errors. The largest comes from the uncertainty of the photon
efficiency. The table also includes correction factors
from the kinematic fit and background contamination.

\begin{table}[h] \centering
\caption{\label{TSystematicError}Summary of correction factors $f_c$ and systematic errors ($\%$).}
\begin{tabular}{l|c|c}  \hline \hline
     & $f_c$ & Sys. err. (\%) \\ \hline
MDC tracking &  & 2.0\\ 
Kinematic fit  & 1.012 & 1.6 \\
Photon efficiency  & & 4.0 \\ 
Backgrounds & 0.984 & 1.6 \\ 
Hadronic model & & 1.6 \\
$B(J/\psi\rightarrow\mu^+\mu^-)$ & & 1.7 \\ 
$B(\pi^0\rightarrow\gamma\gamma)$ & & 0.03 \\ 
MC statistics & & 1.0 \\ \hline
Total & $ 0.996$  & 5.6\\ \hline \hline
\end{tabular}
\end{table}

\subsection{\boldmath Branching fraction}

The branching fraction calculated  with formula (\ref{formula2}) multiplied by the corrrection
factor $f_c$ is

$$B(J/\psi\rightarrow\pi^+\pi^-\pi^0)= (20.91\pm0.21\pm1.16)\times 10^{-3}.$$

\section{Final Result and Discussion}

The absolute branching fraction of $J/\psi\rightarrow\pi^+\pi^-\pi^0$
has been determined using a sample of 58 million $J/\psi$ decays, as
well as by measuring the relative branching fraction of $J/\psi \rt
\pi^+\pi^-\pi^0$ to   $J/\psi \rt \mu^+ \mu^-$ in
$\psi(2S) \rt \pi^+  \pi^- J/\psi$ decays with a sample of 14 million $\psi(2S)$
events. 
The results are in good agreement.
The weighted mean of these two measurements is
$$B(J/\psi\rightarrow\pi^+\pi^-\pi^0)=(2.10\pm0.12)\% .$$

The only reported branching fraction for $\jpsi\ar\ppp$ is by 
Mark-II~\cite{markii},  whereas
many experiments have reported measurements for $\jpsi\ar\rho\pi$
\cite{marki,brau,desy1,pluto,dasp,markiii,besi},
which contributes the dominant part of the $\ppp$ final state.
The result obtained here is higher than those of previous
measurements and has better precision.

\acknowledgments

   The BES collaboration thanks the staff of BEPC for their 
hard efforts. This work is supported in part by the National 
Natural Science Foundation of China under contracts 
Nos. 19991480, 10225524, 10225525, the Chinese Academy
of Sciences under contract No. KJ 95T-03, the 100 Talents 
Program of CAS under Contract Nos. U-11, U-24, U-25, and 
the Knowledge Innovation Project of CAS under Contract 
Nos. U-602, U-34(IHEP); by the National Natural Science
Foundation of China under Contract No. 10175060 (USTC); 
and by the Department of Energy under Contract 
No. DE-FG03-94ER40833 (U Hawaii).


\begin{thebibliography}{120}
\bibitem{rhopi} Generator for generating $\rho\pi$ events 
with the angular distribution of $\jpsi$ decays into 
pseudoscalar plus vector mesons. The angular distribution is described by
\begin{center}
${\frac{d^3\sigma}{d\cos\theta_V d\cos\theta_1 d\phi_1}}
={\sin^2\theta_1[1+\cos^\theta_V+\sin^2\theta_V\cos(2\phi_1)]}$
\end{center}
where $\theta_V$ is the angle between the vector meson and the positron
direction. For $\rho\ar\pi\pi$, $\theta_1$ and $\phi_1$ are the polar and azimuthal angles of the 
$\pi$ with respect to the helicity direction of the $\rho$.
\end{thebibliography}
\end{document}